# S3Mirror: Making Genomic Data Transfers Fast, Reliable, and Observable with DBOS

Steven Vasquez-Grinnell [1], Alex Poliakov [2]

[1] Bristol Myers Squibb
[2] DBOS, Inc.

March 24, 2025

## Summary

To meet the needs of a large pharmaceutical organization, we set out to create S3Mirror - an application for transferring large genomic sequencing datasets between S3 buckets quickly, reliably, and observably. We used the DBOS-Transact durable execution framework to achieve these goals and benchmarked the performance and cost of the application. S3Mirror is an open source DBOS Python application that can run in a variety of environments, including DBOS Cloud Pro where it runs as much as 40x faster than AWS DataSync at a fraction of the cost. Moreover, S3Mirror is resilient to failures and allows for real-time file-wise observability of ongoing and past transfers.

## 1. Background: Computational Genomics in the AWS Cloud

For clinical trials and translational research, biological samples are routinely submitted for sequencing of the genome, epi-genome, and transcriptome. This sequencing is often performed by outside companies who then share the results back with the contracting organization. Specifically - the sequencing provider uploads a batch of raw sequencing reads, in the form of gzipped FASTQ files, to a vendor-specific S3 bucket shared with the company, which must then transfer these files to their own S3 bucket in order to process the raw sequencing data into analysis-ready data and for archiving in the event the data needs to be re-interrogated at a later date. Since these results may be used for portfolio decisions or submission to a health authority such as the FDA or EMEA, the end-to-end process must be traceable and reproducible.

This S3-to-S3 transfer is the focus of our study. The transfer needs to be reliable, observable and efficient. In particular, the following challenges must be met.

## 1.1. Challenge 1: Data Volume

The datasets are large and increasing in size as newer sequencing instruments tend to generate higher volumes and sequencing costs continue to decrease in price. Currently, it is common to generate between 10GB and 100GB of data per sample while sequencing hundreds or even thousands of samples per batch. There may be several batches per week. To transfer large volumes of data, AWS recommends making one concurrent 8-16MB byte range request for each 85–90 MB/s of desired network throughput. [1] One such thread might require over a day of wall-clock time to finish one batch. Thus, the desired solution should leverage many parallel requests to transfer data as quickly as possible.

## 1.2. Challenge 2: Transfer Errors

Any given S3 API call can fail with an intermittent error that is resolved on retry. [2] Other errors require human attention. For example, in one case, some of the files (out of hundreds) did not have appropriate S3 read permissions set. Thus the transfer worked for some files but failed for others, requiring time-consuming work to find all the files affected. Aside from errors like these, the transfer tool itself, or the infrastructure hosting it, can fail or be preempted. In such cases, the naive approach might be to restart the batch from the beginning, which needlessly repeats expensive and time-consuming work. The desired solution should automatically retry to resolve intermittent errors, fail gracefully and with notification on errors that need human intervention, and, if interrupted, have the ability to resume a transfer without repeating completed files.

## 1.3. Challenge 3: Poor Observability

If a transfer takes hours to complete, it is not reasonable to expect a human to monitor its status in real time. And, when transferring large numbers of files using automated, parallelized scripts, it may be difficult for a human observer to notice errors. Furthermore, given a large number of batch deliveries, identifying process failures and the corresponding set of logs for forensic examination can be tedious. If the transfer is somehow prematurely terminated as discussed above, the log of events could even be lost altogether. The desired tool should durably store the file-wise log of all successes and failures and make it observable during and long after the transfer.

# 2. The DBOS Implementation

For large bucket-to-bucket S3 transfers, Amazon recommends the UploadPartCopy API call [3] in which the data is transferred directly in the S3 back plane, without the client having to download and re-upload it. Typically, a large file is split into many byte ranges, around 8-16MB in size, and an UploadPartCopy request is sent for each range. The Amazon `boto3` package provides convenient methods for generating and executing these requests for a given file. It's possible to run many requests for a single file in parallel, and, simultaneously, several files can be transferred at the same time. However, there is a global S3 limit of no more than 3500 simultaneous write requests per bucket prefix. [4] We implemented S3Mirror as a DBOS Transact Python application on top of `boto3`. S3Mirror works as follows.

To start transferring a batch, the client POSTs a payload with the specific bucket and file keys to the application's `/start_transfer` route. This starts an asynchronous DBOS workflow called `transfer_job` whose UUID maps to this specific request. The UUID is immediately returned to the client for tracking purposes.

The core of our architecture is a DBOS Queue - distributed, backed by Postgres and exposed through a lightweight application layer interface. [5] For each file, `transfer_job` puts a DBOS step called `s3_transfer_file` on the transfer queue, keeping a list of Workflow handles [6] to all the enqueued steps. The step `s3_transfer_file` wraps around a `boto3 s3.copy` call configured to transfer one file using some parallelism. This step is decorated to retry up to 3 times with exponential backoff on error.

After enqueueing all the steps, `transfer_job` starts looping over all handles and checking their status. As it iterates, it updates a `tasks` list that tracks the status of each file with statistics like size, transfer time and error, if any. The `transfer_job` then uses `set_event` to persist `tasks` to the database. To view the status of the transfer, the client can retrieve the most recent copy of `tasks` via a GET request to the `/transfer_status/{UUID}` route. The `transfer_job` stops iterating once all tasks run to completion - whether successfully or due to errors.

Currently, DBOS Cloud runs applications using Firecracker MicroVMs with 512MB RAM each by default. We tuned the transfer queue `concurrency` and `worker_concurrency` settings to process multiple files in parallel and enable queue-based auto-scaling in DBOS Pro, while staying below the S3 3500 request limit and keeping individual VMs from running out of RAM.

The application is available in the DBOS Demo Apps repo. [7] Because DBOS is a Python library, S3Mirror can be hosted in a variety of environments with the main requirement being access to a Postgres database.

## 3. Results

After implementing S3Mirror, we tested S3Mirror for reliability and observability. We benchmarked the transfer rates using the publicly available Google Brain Genomics dataset, [8] using the AWS Datasync [9] tool for comparison. Finally, we deployed S3Mirror in our environment and verified the application performance and functionality on several real clinical trial datasets in production.

### 3.1. DBOS Performance Benchmarks

When running S3Mirror locally on a single node, with both buckets in the us-east-1 region, we measured transfer rates around 4 GiB/s. We then repeated the tests in the DBOS Cloud Pro tier which automatically scales to more VMs in response to queue growth.

We artificially increased the size of the Google Brain Genomics dataset, making several copies of each file, to resemble the size of a CG batch. Thus our benchmark dataset was 448 files totaling 11.88TiB in size. We ran this transfer between two buckets located in us-east-1, using s3mirror in DBOS Cloud Pro versus the AWS DataSync tool in Enhanced Mode [10]. In our tests, using the default settings "aws s3 sync" command on this data results in a transfer rate of at most 200 MiB/s and would take over 17 hours to transfer the data. AWS DataSync with enhanced mode ran at the rate of 622.03 MiB/s, taking about 5.6 hours to complete. S3mirror, on the other hand, ran at 24.9 GiB/s taking only 8.1 minutes to finish. These rates are compared in Table 1 below.

| Benchmark Description | Transfer Rate | Transfer time (11.9 TiB) | Comparison |
|---|---|---|---|
| AWS s3 sync command (default) | 0.2 GiB/s | 17+ hours | Basis |
| AWS DataSync (enhanced mode) | 0.6 GiB/s | 5.6 hours | ~ 3x faster |
| s3mirror on a single server | 4.1 GiB/s | 49.5 minutes | ~ 20 x faster |
| s3mirror in DBOS Cloud Pro | 24.9 GiB/s | 8.1 minutes | > 120x faster |

Table 1. S3Mirror intra-region transfer rates compared to the AWS DataSync utility

During this trial, S3mirror auto-scaled to over 10 VMs and accelerated to peak rates above 30 GiB/s. But the transfer only spent a fraction of its 8.1 minute duration at "full speed." Thus larger transfers should see a higher overall rate. We did further tuning experiments and confirmed that we were able to saturate the 3500 simultaneous request S3 limit, at which point it became the speed-limiting bottleneck.

### 3.2. AWS Datasync Cost Comparison

We compared the costs of the transfers in the preceding section. AWS Datasync with Enhanced Mode charges $0.015 per GB transferred, plus $0.55 per task execution. [11] The cost to transfer 11.88TiB is thus $183.03. On the other hand, DBOS Cloud Pro charges $0.05 per 1 million CPU milliseconds. [12] This transfer uses about 2 million for a total of $0.10. These costs are summarized in Table 2.

| Benchmark Description | Cost Calculation | Transfer Cost |
|---|---|---|
| AWS DataSync (Enhanced mode) | 12,165 GB * $0.015 per GB + $0.55 | $ 183.03 |
| s3mirror in DBOS Cloud Pro | 2 Million CPU ms * $0.05 per million | $   0.10 |

Table 2. Comparing the costs of AWS Datasync versus s3mirror in DBOS Cloud Pro

It is important to note that DBOS Cloud Pro requires a $99 per month subscription not shown in the table. Still, even if one were to purchase a month's subscription only for this one transfer, they would pay $99 instead of $183.03 - a savings of nearly 2x.

Both methods incur additional costs for storage and the S3 copy API calls - not listed above.

## 3.3. Reliability and Observability

DBOS Steps execute at least once and do not repeat after successful execution is recorded while Workflows always continue to completion. We ran reliability tests locally, starting a transfer, terminating the application process and re-starting. We ran similar tests in DBOS Cloud by adding a `/crash` POST API handler to the app that immediately called `os._exit(1)` to terminate the process and force DBOS Cloud to recover it.

In all cases we observed interrupted transfers continue to completion without revisiting successfully transferred files. Because we use one DBOS step per file, and because steps have *at least once* guarantees, when restarting after a crash, we observed a re-transfer of the few files that had been mid-flight during the crash. This does not pose a problem because S3 transfers are idempotent in all respects, except for a small storage leak left behind by a partial transfer. Cleaning up such leaks is a simple procedure, already recommended by Amazon as a regular maintenance step when using S3 generally. [13]

As for observability, we found that the `/transfer_status/{UUID}` endpoint provides a convenient way to monitor the transfer. This endpoint supports frequent GET requests without notably affecting transfer performance. At the same time, all the data returned by this endpoint is stored in Postgres. Thus file-wise statistics are available during the transfer and long after it is over.

By comparison AWS Datasync makes the file-wise summary available only after the transfer is complete, persisted as a file in an S3 bucket. We find it much more convenient to have a GET endpoint that we can start polling as soon as the transfer starts. We did not find documentation on whether AWS Datasync would automatically resume intra-S3 transfers after an AWS outage.

## 3.4. Benchmarks on clinical datasets

We deployed S3Mirror and tested it on a few large real-world transfers.

In one case, a Phase 3 glioblastoma trial dataset, having 989 files totaling 8,785 GB in size, transferred at the rate of 3.8 GB/s in about 39 minutes. Transferring the same dataset using the existing AWS CLI method was estimated at over 13 hours.

In a second case, a Phase 3 colorectal cancer trial having 1,056 files totalling 13,289 GB in size, transferred in under 54 minutes, at the rate slightly above 4 GB/s, consistent with the previous trial.

Both cases represent a >20x speed improvement over the previous transfer method using the AWS CLI and enabled the team to deliver urgent results to analysts a full business day faster.

## 4. Discussion

We sought to determine whether the DBOS lightweight durable execution architecture can apply to intra-S3 transfers to meet the challenges of performance, reliability and observability. We have achieved these three objectives. Performance-wise, after some auto-scaling ramp up, we are able to transfer data as quickly as S3 limitations allow. Of course, increasing speed is a matter of adding parallelism. This alone is doable by parallelizing boto3 calls or aws s3 sync commands. But increasing speed *while* adding durability and observability is more challenging without a framework like DBOS.

The durable DBOS Queue abstraction is the centerpiece of our architecture, allowing us to meet the three challenges simultaneously: letting VM workers execute tasks in parallel, durably tracking tasks that need to be completed and making this information observable. To implement this architecture without DBOS, we would need to rely on a separate queue service, incurring additional costs, development and limitations. In DBOS, the queue is stored in Postgres and created with a single line of Python. In fact, the entire application is less than 210 lines of code, making it effective to maintain and update. The observability `/transfer_status/{UUID}` endpoint was particularly useful and will enable the development of a GUI dashboard to display data ingestion status in real-time to non-technical stakeholders awaiting the analysis results.

We realize that the application, in its current form, is tuned specifically for batches having hundreds of files 10-100 GB in size. We did not benchmark s3mirror (or AWS DataSync) on vastly different cases like "millions of 1KB files." S3Mirror may require tuning for optimal performance with those cases. That said, we were pleasantly surprised with both the speed and cost advantage that our approach offers for the case we set out to address.